\documentclass[11pt]{article}
\usepackage{moriond,epsf,amsfonts,latexsym}

\def\Journal#1#2#3#4{{#1} {\bf #2}, #3 (#4)}

\def\PRD{{\em Phys. Rev.} D}

\begin{document}

\centerline{\hfill gr-qc/0306018}
\vspace*{-20pt}

\vspace*{4cm}
\title{SCALAR-TENSOR THEORIES AND COSMOLOGY\\
\&\\
TESTS OF A QUINTESSENCE--GAUSS-BONNET
COUPLING\,\footnote{Contribution to the XXXVIIIth Rencontres de
Moriond on {\it Gravitational Waves and Experimental Gravity},
Les Arcs (France), March 23--29 2003.}}

\author{Gilles ESPOSITO-FARESE}

\address{${\cal G}{\mathbb R}\varepsilon{\mathbb C}{\cal O}$,
Institut d'Astrophysique de Paris,\\ 98bis boulevard Arago, F-75014
Paris, France}

\maketitle\abstracts{
Scalar-tensor theories are the best motivated alternatives to
general relativity and provide a mathematically consistent
framework to test the various observable predictions. They can
involve three functions of the scalar field: (i)~a potential (as
in ``quintessence'' models), (ii)~a matter--scalar coupling function
(as in ``extended quintessence'', where it may also be rewritten
as a nonminimal coupling of the scalar field to the scalar
curvature), and (iii)~a coupling function of the scalar field
to the Gauss-Bonnet topological invariant. We recall the main
experimental constraints on this class of theories, and underline
that solar-system, binary-pulsar, and cosmological observations
give qualitatively different tests. We finally show that the
combination of these data is necessary to constrain the existence
of a scalar--Gauss-Bonnet coupling.}

\section{Introduction}
In the most natural alternative theories to general relativity (GR),
gravity is mediated not only by a (spin-2) graviton corresponding to
a metric $g_{\mu\nu}$, but also by a (spin-0) scalar field $\varphi$.
Such scalar partners generically arise in all extra-dimensional
theories, and notably in string theory. A dilaton is indeed already
present in the supermultiplet of the 10-dimensional graviton, and
several other scalar fields (called the moduli) also appear when
performing a Kaluza-Klein dimensional reduction to our usual
spacetime. They correspond to the components of the metric
tensor $g_{mn}$ in which $m$ and $n$ label extra dimensions. Moreover,
contrary to other alternative theories of gravity, scalar-tensor
theories respect most of GR's symmetries: conservation laws,
constancy of non-gravitational constants, and local Lorentz
invariance even if a subsystem is influenced by external masses. They
can also satisfy exactly the weak equivalence principle (universality
of free fall of laboratory-size objects) even for a massless scalar
field.

Scalar fields are also involved in the cosmological models which
reproduce most consistently present observational data. In
particular, inflation theory is based on the presence of a scalar
$\varphi$ in a potential $V(\varphi)$ (for instance parabolic).
It behaves as a fluid with a positive energy density $8\pi G
\rho_\varphi = \dot\varphi^2 + 2 V(\varphi)$ but a negative
pressure $8\pi G p_\varphi = \dot\varphi^2 - 2 V(\varphi)$.
This causes a period of exponential expansion of the universe,
which can explain why causally disconnected regions at present
may have been connected long ago. The isotropy of the observed
cosmic microwave background (CMB) can thus be understood.
Inflation also predicts that our universe is almost spatially
flat, just because any initial curvature has been exponentially
reduced by the expansion. This is in remarkable agreement with
the location of the first acoustic peak of the CMB
spectrum at a multipolar index~\cite{wmap03} $\ell \simeq 220$.
Observations of type Ia supernovae~\cite{Perl,Garn} tell us that there
is about 70\% of negative-pressure dark energy in our present universe
($\Omega_\Lambda \simeq 0.7$), suggesting that its expansion has been
re-accelerating recently (since redshifts $z \sim 1$). This can be
explained by the presence of a cosmological constant $\Lambda$ in GR,
but the quantity $\Omega_\Lambda \simeq 0.7$ translated in natural
units gives an extremely small value $\Lambda \simeq 3\times 10^{-122}
c^3/(\hbar G)$, very problematic for particle physics if $\Lambda$ is
to be interpreted as the vacuum energy. This is the main reason why
``quintessence'' models have been proposed, in which the cosmological
constant is replaced again by the potential $V(\varphi)$ of a scalar
field. Its evolution towards a minimum of $V$ during the
cosmological expansion then explains more naturally why the present
value $V(\varphi_0) \simeq \Lambda/2$ is so small.

Besides these theoretical and experimental reasons for studying
scalar-tensor theories of gravity, one of their greatest interests
is to embed GR within a class of mathematically
consistent alternatives, in order to understand better which
theoretical features have been experimentally tested, and which
can be tested further.

The following action defines the most general theory satisfying the
weak equivalence principle and involving only one spin-0 degree of
freedom besides the usual (spin-2) graviton:
\begin{eqnarray}
S&=&S_{\rm matter}[{\rm matter} ; \widetilde g_{\mu\nu} \equiv
A^2(\varphi) g_{\mu\nu}]\nonumber\\
&&+ {c^3\over 4 \pi
G}\int\sqrt{-g}\left\{{R\over 4}-{1\over 2}(\partial_\mu\varphi)^2 -
V(\varphi)\right\}\nonumber\\
&&-\hbar\int\sqrt{-g}~W(\varphi)\left(R_{\mu\nu\rho\sigma}^2
-4R_{\mu\nu}^2+R^2\right)\ ,
\label{action}
\end{eqnarray}
where $\widetilde g_{\mu\nu}$ is the metric to which matter is
universally coupled, and $g_{\mu\nu}$ is the Einstein metric
(describing the spin-2 degree of freedom). This action involves three
function of the scalar field: a coupling function $A(\varphi)$ to
matter, a potential $V(\varphi)$, and a coupling function
$W(\varphi)$ to the Gauss-Bonnet topological invariant. Any other
combination of the curvature tensor would introduce an extra scalar
field in the theory, and/or a second negative-energy massive graviton
which would make the model unstable.

In Sections 2 and 3, we will not consider any scalar--Gauss-Bonnet
coupling, and set $W(\varphi) = 0$. The potential $V(\varphi)$ will
also be neglected in Section 2, in which we will review the main
experimental constraints on the coupling function $A(\varphi)$,
coming from solar-system and binary-pulsar data.\cite{def7,def9}
In Section 3, we will summarize our results concerning the
reconstruction of $A(\varphi)$ and $V(\varphi)$ from cosmological
observations.\cite{beps00,efp01} Section 4 will be devoted to the
scalar--Gauss-Bonnet coupling $W(\varphi)$, which can be constrained
only if one takes into account both solar-system and cosmological
data.\cite{efs03} We will finally give our conclusions in Section 5.

\section{Solar-system and binary-pulsar constraints}
The effects of a massive scalar field has a negligible effect on the
motion of celestial bodies if its mass is large with respect to the
inverse of the interbody distances. On the other hand, if its mass
is small enough, its potential $V(\varphi)$ can be locally neglected,
but its coupling function to matter, $A(\varphi)$, is strongly
constrained by experiment.

\begin{figure}
\centerline{\epsfbox{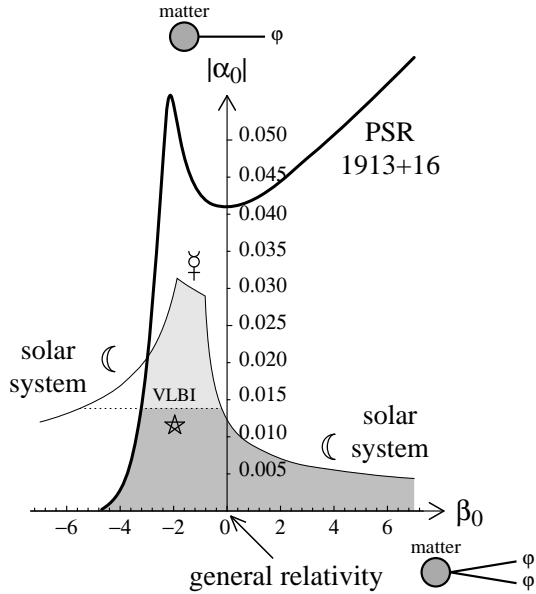}}
\caption{Solar-system and binary-pulsar constraints on
the matter-scalar coupling function $\ln A(\varphi) =
\alpha_0(\varphi-\varphi_0) + {1\over 2} \beta_0
(\varphi-\varphi_0)^2 + O(\varphi-\varphi_0)^3$. The allowed region
is shaded. The vertical axis ($\beta_0=0$) corresponds to
Brans-Dicke theory with a parameter $2\omega_{\rm BD}+3
= 1/\alpha_0^2$. The horizontal axis ($\alpha_0=0$) corresponds to
theories which are perturbatively equivalent to GR, i.e., which
predict strictly no deviation from it (at any order $1/c^n$) in the
weak-field conditions of the solar system.
\label{fig1}}
\end{figure}

The predictions of metric theories of gravity in weak-field
conditions can be parametrized by a set of 10 real numbers in the so
called ``PPN'' formalism (parametrized post-Newtonian). All of them
are presently constrained to be very close to their general
relativistic values, and in particular the two famous Eddington
parameters $\beta$ and $\gamma$ (both equal to 1 in GR). In
scalar-tensor theories,\cite{def7,def9} they are related to the
first two derivatives of $\ln A(\varphi)$, computed at the background
value $\varphi_0$ of the scalar field. They give the constraints
displayed as a thin line in Figure \ref{fig1} (where the Moon symbol
refers to Lunar Laser Ranging, the Mercury symbol to the perihelion
shift of this planet, and the star symbol to light deflection as
measured by Very Long Baseline Interferometry). Solar-system tests
thus constrain the first derivative $\alpha_0
\equiv \partial\ln A(\varphi)/\partial\varphi$ to be small, but do not
tell us much about the second derivative $\beta_0\equiv \partial^2\ln
A(\varphi)/\partial\varphi^2$. If $\alpha_0$ is small enough,
arbitrary large positive or negative values of $\beta_0$ are a priori
allowed.

Binary-pulsar give qualitatively different constraints because of
nonperturbative strong-field effects. Indeed, the largest deviations
{}from the flat metric, at the surface of a star, are of order
$GM/Rc^2\simeq 0.2$ for a pulsar (neutron star), as compared to
$2\times 10^{-6}$ for the Sun. We showed~\cite{def7,def9} that if
$\beta_0$ is negative, then it is energetically favorable for a
neutron star, above a critical mass, to create a nonvanishing scalar
field. Since this is analogous to the spontaneous magnetization of
ferromagnets, we called this effect ``spontaneous scalarization''.
Such macroscopic scalar charges change drastically the physics of a
binary system, notably because it emits dipolar gravitational
(scalar) waves $\propto 1/c^3$, much larger that the usual
quadrupolar radiation $\propto 1/c^5$ predicted by GR. This is the
reason why binary-pulsar data, which are consistent with GR, rule
out scalar-tensor models such that $\beta_0 < -5$, even for
vanishingly small values of $\alpha_0$ (i.e., even if they are
strictly indistinguishable from GR in the solar system).

We also showed that the LIGO/VIRGO interferometers will be more
sensitive to $\beta_0$ than solar-system tests, but binary-pulsar
data are so precise that they already exclude the models which
predict significant effects in the gravitational waveforms. This is
a good news, since it proves that pure GR wave templates suffice to
analyze future LIGO/VIRGO data. On the other hand, it was
shown~\cite{will02} that the LISA interferometer can be sensitive to
scalar effects which are still allowed by all present tests.

In conclusion, solar-system tests tightly constrain the first
derivative of $\ln A(\varphi)$ (linear matter-scalar coupling
strength), whereas binary-pulsar data impose that its second
derivative (quadratic coupling matter-scalar-scalar) is not large
and negative. We will now see that cosmological observations give
access to the full shape of this coupling function, of course not
with the same accuracy as the above tests, but with the capability
of constraining any higher derivative of $\ln A(\varphi)$ (vertex of
matter with any number of scalar lines). Moreover, cosmological data
can also give access to the full shape of the potential $V(\varphi)$.

\section{Reconstruction of a scalar-tensor theory from
cosmological observations}
In cosmology, the usual approach to study quintessence models
is to assume a particular form for the potential $V(\varphi)$
(and the matter-scalar coupling function $A(\varphi)$ when
one considers ``extended quintessence'' models), to compute
all possible observable predictions, and to compare them to
experimental data.

In contrast, in the phenomenological approach, one wishes to
{\it reconstruct\/} the Lagrangian of the theory from cosmological
observations. We proved~\cite{beps00} that the knowledge of the
luminosity distance $D_L(z)$ and of the density fluctuations
$\delta_m(z) = \delta\rho/\rho$ as functions of the redshift $z$
indeed suffices to reconstruct both the potential $V(\varphi)$ and
the coupling function $A(\varphi)$. Although the explicit
reconstruction needs some algebra, this result seems anyway
obvious: It is possible to {\it fit\/} two observed functions
[$D_L(z)$ and $\delta_m(z)$] thanks to two unknown ones [$V(\varphi)$
and $A(\varphi)$].

However, future experiments (like the SNAP satellite) will
only give access to the luminosity distance $D_L(z)$ with
a good accuracy, and the density contrast $\delta_m(z)$
cannot yet be used to constrain the models. A
{\it semi-phenomenological\/} approach can thus be useful: We make
some theoretical hypotheses on either the potential $V(\varphi)$ or
the coupling function $A(\varphi)$, and we reconstruct the other one
{}from $D_L(z)$. A priori, one may think that such a reconstruction is
again obvious: We fit one observed function [$D_L(z)$] with one
unknown function [$V(\varphi)$ or $A(\varphi)$]. But this naive
reasoning is only valid locally, on a small interval. Indeed,
the reconstructed function may for instance diverge for some value
of the redshift, or one of the degrees of freedom may need to
take a negative energy beyond a given redshift, which would make
the theory unstable (and ill defined as a field theory on the
surface where the energy changes its sign). The positivity of the
graviton energy implies $A^2(\varphi) > 0$, which can be translated in
terms of the standard Brans-Dicke scalar field as $\Phi_{\rm BD} > 0$.
On the other hand, the positivity of the scalar-field energy imposes
the minus sign in front of the scalar kinetic term
$-(\partial_\mu\varphi)^2$ in action (\ref{action}), which
translates as $\omega_{\rm BD} > -{3\over 2}$ in terms of the
standard Brans-Dicke parameter. We showed that these conditions impose
tight constraints on the theories as soon as one knows $D_L(z)$ over
a wide enough interval $z\in[0,\sim 2]$.

For instance, we proved that the present accelerated expansion
of the universe can be perfectly described by a scalar-tensor theory
with a vanishing potential $V(\varphi) = 0$ (and therefore a
vanishing cosmological constant too). We derived analytically
the coupling function $A(\varphi)$ which reproduces exactly the same
evolution of the scale factor $a(z)$ as the one predicted by
GR plus a cosmological constant. We even found that the
reconstructed function $\ln A(\varphi)$ has a nice parabolic shape,
with a minimum very close to the present value $\varphi_0$ of the
scalar field, and a positive second derivative. This is not only
consistent with binary-pulsar data (which forbid large and negative
values of this second derivative) but also with the cosmological
attractor phenomenon analyzed by Damour and Nordtvedt~\cite{dn93}:
The scalar field is generically attracted towards a minimum of $\ln
A(\varphi)$ during the cosmological expansion, whereas some fine
tuning would be necessary to reach a maximum (negative second
derivative). Therefore, we are in the difficult situation in which
two very different theories are both consistent with experimental
data, and there seems to be no way to distinguish them. Fortunately,
we found that this scalar-tensor theory cannot mimic GR plus a
cosmological constant beyond a redshift $z \sim 0.7$, because the
scalar field $\varphi$ would diverge at this value, and above all
because the graviton energy would become negative beyond. Therefore,
it suffices to measure $D_L(z)$ precisely enough up to $z \sim 1$ to
rule out such a potential-free scalar-tensor theory. Actually, if
$D_L(z)$ is measured over a wider interval $z\in[0,\sim 2]$, we
showed that large experimental errors (tens of percent) were not
problematic: It is still possible to distinguish this potential-free
model from GR plus a cosmological constant, and thereby to rule out
one of them. The results of the SNAP satellite up to $z \sim 2$ will
therefore be very useful to constrain scalar-tensor theories of
gravity.

One can also impose a particular form of the coupling function
$A(\varphi)$ and reconstruct the potential $V(\varphi)$ which
reproduces the observed luminosity distance $D_L(z)$. For instance,
for a minimally coupled scalar field $A(\varphi) = 1$ (usual
``quintessence'') in a spatially curved universe, we analytically
derived the expression of $V(\varphi)$ which gives the same
cosmological evolution as GR plus a cosmological constant in
a spatially flat universe. We found that the shape of the potential
is smoother when the universe is (marginally) closed. If it is flat
or almost flat, one obviously recovers a cosmological constant with
its unnaturally small value $\Lambda \simeq 3\times 10^{-122}
c^3/(\hbar G)$. Therefore, in that case, aesthetic reasons may help
us discriminate between the theories, instead of the much stronger
argument of the positivity of energy that we used above. This shows
anyway that the sole knowledge of $D_L(z)$ suffices to constrain
scalar-tensor theories of gravity.

\section{Experimental constraints on a scalar--Gauss-Bonnet
coupling}
In order to illustrate the different kinds of experimental
constraints that can be imposed~\cite{efs03} on the
scalar--Gauss-Bonnet coupling function $W(\varphi)$, we will now
focus of a theory with $A(\varphi) = 1$ and $V(\varphi) = 0$ in
action~(\ref{action}).

Solar-system and binary-pulsar tests are local, and any deviation
{}from GR depends on the magnitude of the scalar field created by a
massive body. Let us thus analyze first the equation satisfied by
$\varphi$ in the vicinity of a spherical mass $M_\odot$. We can
assume that the metric is close to the Schwarzschild solution, and
we get at the first nonvanishing order in powers of $GM_\odot/c^2$
\begin{equation}
\Box \varphi = {3 r_0^2\over r^6}\left({2 G M_\odot\over c^2}\right)^2
\left[W'_0 + W''_0 \varphi + O(\varphi^2)\right]\ ,
\label{boxphi}
\end{equation}
where we have set $r_0^2 \equiv 16 \pi G \hbar / c^3$, and where the
derivative $W'(\varphi)$ has been expanded in powers of $\varphi$ in
the right-hand side. Since we are assuming that $\varphi$ takes small
values, let us neglect the contribution $W''_0 \varphi$. We can then
compute any observable prediction, but we quote below only the
results for the light deflection angle ($\Delta\theta_*$) and for the
perihelion shift per orbit ($\Delta\theta_p$), which suffice for our
purpose:
\begin{eqnarray}
\Delta\theta_* &=& {4 G M_\odot\over \rho_0 c^2} +
{1536\over 35}\left({G M_\odot\over \rho_0 c^2}\right)^3
\left({r_0\over \rho_0}\right)^4 W'^2_0\ ,
\label{defl0}\\
\Delta\theta_p &=& {6\pi G M_\odot\over p c^2} +
192\pi\left({G M_\odot\over p c^2}\right)^2
\left({r_0\over p}\right)^4 W'^2_0\ ,
\label{peri0}
\end{eqnarray}
where $\rho_0$ is the minimal distance between the light ray and the
Sun, and $p$ is the {\it semilatus rectum\/} of an orbit. The first
terms on the right-hand sides are the usual general relativistic
predictions, at first order in $G M_\odot/c^2$. In conclusion,
solar-system (and binary-pulsar) tests can easily be passed if
$|W'_0|$ is small enough.

One can now reconstruct the full shape of $W(\varphi)$ from the
cosmological observation of the luminosity distance $D_L(z)$, as in
the previous section. We found that this can always
been done, without any problem of negative energy, contrary to what
we saw in Section 3. Moreover, there exists again an attraction
mechanism which drives the scalar field towards a minimum of
$W(\varphi)$ during the cosmological expansion. Therefore, a small
value of the slope $|W'_0|$ is indeed expected at present,
consistently with what is needed for solar-system tests. In
conclusion, we are faced again with a serious problem: We just found
a theory which seems to be consistent with all experimental data,
although it is very different from GR in its field content. Our aim
is therefore to find a way to distinguish it from GR, or to rule it
out for internal consistency reasons.

\begin{figure}
\centerline{\epsfbox{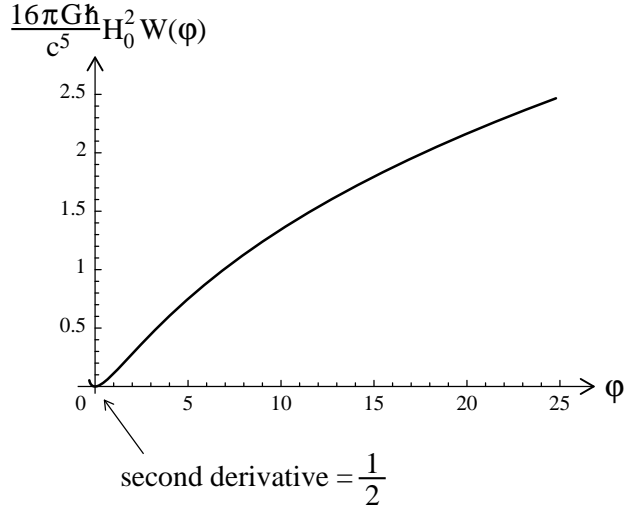}}
\caption{Scalar--Gauss-Bonnet coupling function $W(\varphi)$ which
exactly reproduces the cosmological expansion predicted by GR plus a
cosmological constant.
\label{fig2}}
\end{figure}

Figure \ref{fig2} displays this reconstructed coupling function
$W(\varphi)$, in which the present value of the scalar field is close
to the minimum at $\varphi=0$. Its shape is nicely smooth, but its
second derivative at the origin is huge if one divides it by the
tiny factor $(16\pi G\hbar/c^5) H_0^2$, where $H_0$ denotes the
Hubble constant. One gets $W''_0 \simeq 7\times 10^{119}$, which is in
fact not surprizing, since the coupling function $W(\varphi)$ behaves
in action (\ref{action}) as the inverse of a cosmological constant.
Indeed, $W(\varphi)$ multiplies the square of the curvature tensor,
whereas the usual Einstein-Hilbert term involves the first power of
the scalar curvature, and a cosmological constant does not multiply
any curvature term at all. Therefore, it was expected that
$W(\varphi)$ involve a dimensionless number of the order of the
inverse of $(\hbar G/c^3)\Lambda \simeq 3\times 10^{-122}$.
Therefore, this model is ugly, but it is not yet ruled out. One
should not confuse fine tuning and large (or small) dimensionless
numbers in a model. We are here in the second situation, but there is
a priori no fine tuning since the scalar field is attracted towards
the minimum of $W(\varphi)$ during the cosmological expansion. There
remains to study how efficiently it is attracted, but this is
actually not necessary for our purpose.

Indeed, $W''_0$ takes such a gigantic value that an approximation
that we made to analyze solar-system tests is no longer valid.
Indeed, we have $|W''_0\varphi| \gg |W'_0|$, so that the second
term on the right-hand side of Eq.~(\ref{boxphi}) cannot be
neglected. To simplify the discussion, we will anyway assume that
$W(\varphi)$ is parabolic, which is a good approximation in a
vicinity of the minimum $\varphi = 0$. We will thus neglect the
higher order terms $O(\varphi^2)$ in Eq.~(\ref{boxphi}); taking them
into account would not change our conclusions below. We did not find
a close analytic solution to Eq.~(\ref{boxphi}), but is is possible
to write it as a series
\begin{eqnarray}
\varphi&=&{W'_0\over W''_0}
\sum_{n\geq 1}{1\over (3\times 4)(7\times 8)\cdots (4n-1)(4n)}
\left({12 r_0^2 G^2 M_\odot^2 W''_0\over r^4 c^4}\right)^n
\label{phiexact}\\
&\simeq&{W'_0\over W''_0}
\left[{\cos\atop \cosh}\left({G M_\odot r_0\over r^2
c^2}
\sqrt{3|W''_0|}\right)-1\right]\qquad
{\hbox{if $W''_0<0$,}\atop\hbox{if $W''_0>0$.}}
\label{phiapprox}
\end{eqnarray}
The second expression is a good approximation if the argument of
the cosine (or hyperbolic cosine if $W''_0 > 0$) is much greater than
1. This is the case if we use the huge value of $W''_0$ obtained
above from the cosmological reconstruction, and a typical solar-system
distance for the radius $r$: The argument of the hyperbolic cosine
is then of order $10^8$.

The above solution is such that $\varphi \propto W'_0$, therefore
we do not find any nonperturbative effect similar to the
``spontaneous scalarization'' of neutron stars mentioned in Section 2
above. Moreover, $\varphi \rightarrow 0$ as $ r \rightarrow \infty$,
and we recover GR for distances $r > 4\times 10^{14}~\rm{m}$
(i.e., farther that the solar system including Oort's comet cloud).
On the other hand, there are highly nonlinear corrections proportional
to $1/r^{4n}$ within the solar system. Since the ratio $(12 r_0^2 G^2
M_\odot^2 W''_0/ r^4 c^4)$ is much greater that $1$, its successive
powers blow up, but they are compensated by the factors
$1/(3\times4\times7\times\cdots 4n)$ which behave like the inverse of
factorials. Therefore, the successive terms of series (\ref{phiexact})
start to grow exponentially, then reach a maximum for a value of the
index $n$ which may be large, and finally tend towards zero. Each of
these successive terms must be assumed to be small enough for the
model to pass all classical tests, but one should not forget that the
largest one does not correspond to $n=1$.

In order to study the effects of such highly nonlinear terms in the
solar system, we compute their corrections to the Schwarzschild
metric in the form
\begin{equation}
ds^2 = -\left(1+\sum_{n\geq 1}{\beta_n\over \rho^n}\right)c^2 dt^2 +
\left(1+\sum_{n\geq 1}{\alpha_n\over \rho^n}\right)d\rho^2 +\rho^2
\left(d\theta^2 +\sin^2\theta\, d\phi^2\right)\ ,
\end{equation}
and we find that the light deflection angle and the perihelion shift
are respectively given by
\begin{eqnarray}
\Delta\theta_* &=& \sum_{n\geq 1} 2^{n-1}{\Gamma\left({n+1\over
2}\right)^2\over
\Gamma(n+1)}\,{\alpha_n-n\beta_n\over\rho_0^n} +
O(\alpha_n,\beta_n)^2\ ,
\label{defl}\\
\Delta\theta_p &=& {6\pi G M_\odot\over p c^2} -
\sum_{n\geq 3} {n(n-1)\beta_n c^2\over 2 GM_\odot p^{n-1}}\,\pi +
O(\alpha_n,\beta_n)^2\ ,
\label{peri}
\end{eqnarray}
which generalize Eqs.~(\ref{defl0})-(\ref{peri0}) above. Note that
these results are (independently) perturbative in each of the
coefficients $\alpha_n$ and $\beta_n$, because we are assuming that
the scalar-field effects are negligible with respect to the general
relativistic predictions. However, the dominant scalar-field
corrections may correspond to a large value of index $n$.

When solution (\ref{phiexact}) or its approximation
(\ref{phiapprox}) are used to compute the metric coefficients
$\alpha_n$ and $\beta_n$ in the above observable predictions,
and if we use the huge value of $W''_0$ obtained from the previous
cosmological reconstruction of $W(\varphi)$, we get the following
experimental constraint:
\begin{equation}
|W'_0| < 10^{-2\times 10^{11}}.
\end{equation}
Now we can speak of fine tuning, and even of {\it hyperfine\/}
tuning! This constraint simply means that the present value of the
scalar field must be {\it exactly\/} at the minimum of the coupling
function $W(\varphi)$, otherwise solar-system tests are violated. And
since the universe is still evolving, the scalar field cannot remain
so close to the minimum for more than a fraction of a second.
Therefore, even if we assumed that $W'_0 = 0$ strictly to pass
solar-system tests, this would not be the case a tiny instant later.
In conclusion, we managed to rule out the scalar-tensor model
$A(\varphi) = 1$, $V(\varphi) = 0$ and $W(\varphi) \neq 0$. It
cannot describe an accelerating expansion of the universe at present
and pass solar-system (and binary-pulsar) tests at the same time.

Of course, this result does not rule out any scalar--Gauss-Bonnet
coupling. A model with three (or even two) free functions
$A(\varphi)$, $V(\varphi)$ and $W(\varphi)$ can obviously pass all
present tests. For instance, GR plus a cosmological constant simply
corresponds to $A(\varphi)=1$, $V(\varphi)=\Lambda/2$ and
$W(\varphi)=0$. But the presence of a non-constant coupling
$W(\varphi)$ can change the physics at small scales, notably in the
very early universe (Big-Bang) and for later clustering properties.

The fact that $W(\varphi)$ induces effects at small scales can be
understood by a simple dimensional argument. Since this
function multiplies the square of the curvature in action
(\ref{action}), it induces corrections proportional to $1/r^7$ (and
higher orders) to the Newtonian potential in $1/r$, and thereby
generically dominates at small scales. However, we saw above that this
quick reasoning can be erroneous in some perturbative but highly
nonlinear situations. Indeed, if $W''_0$ takes very large and {\it
negative\/} values, the cosine involved in Eq.~(\ref{phiapprox})
shows that $\varphi$ is always of the order of $-W'_0/W''_0$, even for
small distances $r$. One can then prove that the (very easily
satisfied) condition $|r_0^2 W'_0| \ll r^2$ suffices for all
scalar-field effects to be negligible in the solar system, even if
$|W''_0| \sim 10^{120}$. This remark underlines that nonlinear effects
can drastically change the intuitive behavior, but let us recall
that our cosmological reconstruction above predicted a large and {\it
positive\/} value for $W''_0$. In that case, we did find that
the scalar--Gauss-Bonnet coupling induces large effects at small
scales, and even exponentially larger than the linear results
(\ref{defl0})-(\ref{peri0}).

\section{Conclusions}
Scalar-tensor theories of gravity are the best motivated
alternatives to general relativity. Three classes of experimental
data give {\it qualitatively\/} different constraints on them.
Solar-system tests strongly constrain the first derivative of the
matter-scalar coupling function $A(\varphi)$ (i.e., the linear
matter-scalar coupling strength). Binary-pulsar data forbid large and
negative values of its second derivative (quadratic
matter-scalar-scalar coupling). The knowledge of the two cosmological
functions $D_L(z)$ and $\delta_m(z)$ suffices to reconstruct the full
shape of both $A(\varphi)$ and the potential $V(\varphi)$ on a
finite interval of $\varphi$. The knowledge of the luminosity distance
$D_L(z)$ alone over a {\it wide\/} redshift interval strongly
constrains the theories if one takes into account solar-system (and
binary-pulsar) data, the positivity of the graviton and scalar
energies, and the stability and naturalness of the models. Future
data, provided by experiments like the SNAP satellite, will
notably allow us to discriminate between GR plus a cosmological
constant and a potential-free scalar-tensor theory. The possible
coupling $W(\varphi)$ of the scalar field to the Gauss-Bonnet
topological invariant can be constrained only if one takes into
account cosmological and solar-system data together. The predictions
of the model at small distances can depend on highly nonlinear
corrections. Of course, a model including all three functions
$A(\varphi)$, $V(\varphi)$ and $W(\varphi)$ is experimentally
allowed, since GR plus a cosmological constant is a particular case.
The presence of a scalar--Gauss-Bonnet coupling $W(\varphi)$ will
generically change the behavior of the theory at small scales
(clustering, Big Bang).

\end{document}